%% file: main.tex
\documentclass[reqno]{amsart}
\usepackage{amsmath}
\usepackage{amssymb}
\usepackage{amsthm}
\usepackage{leftidx}
\usepackage{tikz-cd}
\usepackage{ textcomp }
\usepackage{booktabs}
\usepackage{dblfloatfix}

\newtheorem{defx}{Definition}[section]

\newcommand{\cost}{\text{Cost}}

\title{BitCoin Coin Selection with Leverage}
\author{Daniel J. Diroff, Akvelon Inc.}

\begin{document}

\begin{abstract}
    We present a new Bitcoin coin selection algorithm, ``coin selection with leverage", which aims to improve upon cost savings than that of standard knapsack like approaches. Parameters to the new algorithm are available to be tuned at the users discretion to address other goals of coin selection. Our approach naturally fits as a replacement for the standard knapsack ingredient of full coin selection procedures.
\end{abstract}

\maketitle

\section{Introduction}

A common problem facing Bitcoin wallet providers and exchanges is that of coin selection, selecting Unspent Transaction Outputs (UTXOs) from their internal collection to fund customer payment requests. In this context, there are many contradictory goals that one usually has in mind. A great breakdown of these issues is available in the master's thesis of M. Erhardt \cite{erhardt_2016}.

A Bitcoin transaction is the record of an exchange of Bitcoin. It contains information on the input funds (UTXOs) as well as where these funds are to be sent (output funds). As incentive for a Bitcoin miner to pick up a transaction, a transaction fee is attached in an implicit form. The sum total of the input UTXOs will cover more than the sum of the outputs, it is understood that what remains is the fee. In this sense, it is the sender which has control over the transaction fee to be paid. Today miners usually prioritize by fee-per-byte, and there is a market for a typical fee-per-byte rate for certain expected confirmed times. For simplicity, in our analysis we assume we have this market fee-per-byte rate given, which we will denote by $\gamma$ (the various different rates for different expected confirmed times we will not address).

Due to the Bitcoin protocol, an individual UTXO can ``only be spent in its entirety". Because of this, it is frequently the case that an additional output is included, normally referred to as a \emph{change output}, which sends the excess funds back to the sender. The presence of this change output can be avoided if the total value of the input UTXO is exactly enough to cover the outputs and the transaction fee.

Moreover, transactions lacking a change output (a \emph{change-free} transaction) are smaller in byte size, hence a lower transaction fee is needed for the same expected confirmation time. This goal of cost minimization is (one of) the main goal(s) behind coin selection, in addition to others such as user privacy and addressing the rampant growth of the UTXO set \cite{erhardt_2016}, \cite{maxwell}, \cite{lopp_2015}. It will be the main focus of this paper.

This coin selection problem naturally can be seen as a ``knapsack type" problem where, more-or-less, a standard knapsack algorithm can be implemented. In fact, a similar algorithm is currently being used for coin selection in Bitcoin Core, but as mentioned above, there are other goals in mind rather than just minimizing cost. 

In this work we propose a technique which we call \emph{coin selection with leverage} which aims to improve on the standard knapsack technique and to address the goal of minimizing cost. Furthermore, the technique will have many parameters that can be tuned at the user's discretion to address the above mentioned additional goals for coin selection.

In Section \ref{prob_statement} we give the precise mathematical formulation of the problem as well as the setup. In Section \ref{full_soln} we explain both the new leverage technique and the standard one for which we compare. Section \ref{sim_results} includes simulation results done to compare the two approaches. The code is publicly available on GitHub \cite{GitHub}.

\section{Statement of Problem and Basic Solutions}
\label{prob_statement}

Here we formulate the problem precisely so as to set up a proper comparison of our leverage technique with that of the standard knapsack approach. 

Before we give the statement of the problem, there are some relevant constants that we use below (we assume all transactions are the P2PKH format):
\begin{enumerate}
    \item $\gamma$: The market rate for the transaction fee-per-byte
    \item $10$: Number of bytes required for metadata/overhead in a transaction
    \item $148$: Number of bytes required to record each input in a transaction
    \item $34$: Number of bytes required to record each output in a transaction
    \item $d$: The Bitcoin dust threshold, typically $d = (148 + 34)\gamma$ Satoshi
    \item $h$: The ``make change" threshold. Maximal overpayment amount (further explained below).
\end{enumerate}

The problem is somewhat elaborate, we formulate it as follows: We assume we have:

\begin{enumerate}
    \item A finite sequence $\mathcal{U}$ of $n$ UTXO's available to the wallet provider. We will typically denote this by:
    $$
    \mathcal{U} = \{ u_1, u_2, \dots, u_n \},
    $$
    and assume these are ordered in a decreasing fashion, i.e. $u_1 \geq u_2 \geq \cdots u_n$. (More precisely $\mathcal{U}$ is an $n$-tuple in $\mathbb{R}^n$).
    \item A finite sequence $\mathcal{P}$ of $m$ payment requests which are needed to be processed by the wallet provider. We denote these by:
    $$
    \mathcal{P} = \{ p_1, p_2, \dots, p_m \}.
    $$
    By assumption we take $\mathcal{P}$ to be \emph{ordered by urgency}, i.e. the urgency for the gateway to process payment $p_k$ is greater than that of $p_j$ if $k < j$. This notion of urgency is not crucial for what comes below, but it will help in simplifying notation and aid in intuition.
\end{enumerate}

We break up the problem into two sub problems which we call \emph{basic} and \emph{full}. Before stating what these are we make the following definitions:

\begin{defx} \label{transdef}
A transaction is a $4-$tuple $(I,J,c,r)$ where $I \subset \mathcal{U}$, $J \subset \mathcal{P}$ are subsequences and $c,r \in \mathbb{R}_{\geq 0}$. We say:
\begin{enumerate}
    \item The $\textbf{size}$ (in bytes) of the transaction is
    $$
    \emph{Size}(I,J,c,r) := 10 + 148|I| + 34|J| + 34(1-\delta_{c,0}),
    $$
    where $\delta_{c,0}$ is the kronecker delta (i.e. $1$ if $c=0$ and $0$ otherwise).
    For convenience, we will also make use of the function \emph{size}, which is simply the affine linear function
    $$
    \emph{size}(m,n,a) = 10 + 148n + 34m +34a.
    $$
    These functions are related by:
    $$
    \emph{Size}(I,J,c,r) = \emph{size}(|I|, |J|, 1-\delta_{c,0}).
    $$
    \item The transaction is $\textbf{change-free}$ if $c=0$.
    \item The transaction is $\textbf{valid}$ if
    $$
    \sum_{u \in I} u \geq \sum_{p \in J} p + \emph{Size}(I,J,c,r)\gamma
    $$
    and
    $$
    \sum_{u \in I} u + c + r = \sum_{p \in J} p + \emph{Size}(I,J,c,r)\gamma
    $$
    \item \label{goodtrans} The transaction is $\textbf{good}$ if it is valid and either one of the following two cases hold:
    \begin{enumerate}
        \item $c=0$ and $0 \leq r \leq h$ ($h$ is the make-change threshold from above), or
        \item $r=0$ and $c \geq d$.
    \end{enumerate}
    \item \label{costtrans} The $\textbf{cost}$ of the transaction is
    $$
    \emph{Cost}(I,J,c,r) := \emph{Size}(I,J,c,r)\gamma + r
    $$
    (The units here are usually in Satoshi, i.e. consistent with the units of $\gamma$.)
\end{enumerate}
\end{defx}

One thinks of a transaction $(I,J,c,r)$ as $I$ representing the input UTXOs, $J$ representing the desired outputs (or payment requests), $c$ being the change output amount (which could be zero) and lastly $r$ giving, what one might call, the overpayment amount, i.e. the small extra amount given to the miner as extra incentive to pick up the transaction. Note for $r$, it is desired that it be less than the make-change threshold $h$ which is usually taken to be less than the dust threshold $d$, otherwise it reasoned that it should be returned to the sender in the form of change. This is reasoning behind Definition \ref{transdef}.(\ref{goodtrans}).

By a \emph{basic} problem we mean finding a good transaction to process a fixed collection of payment requests. That is, given some specified $J \subset \mathcal{P}$ collection of payment requests, we look for $I \subset \mathcal{U}$, $c, r \in \mathbb{R}_{\geq 0}$ so that $(I, J, c, r)$ is a good transaction.

By the \emph{full} problem we mean for some fixed sub collection of payment requests $\mathcal{P}' \subset \mathcal{P}$, finding a collection of good transactions $T_1 = (I_1, J_1, c_1, r_1), \dots, T_K = (I_K, J_K, c_K, r_K)$ that as a whole will process \emph{at least} all of $\mathcal{P}'$, i.e. $\cup J_k \supseteq \mathcal{P}'$ with the $J_k$ pairwise disjoint.

We present below three algorithms which solve the basic problem, each of which solves it with some slightly different goals in mind desirable by the BTC wallet provider. These three algorithms are then combined in two different ways to create algorithms to solve the full problem. We call these procedures \emph{Knapsack Coin Selection} and \emph{Knapsack Coin Selection with Leverage}, the former of which is similar to parts of what can be found being utilized today. The latter is the new technique and we show below through simulations it's utility.

\subsection{Basic Problem - Fallback Solution}
The first algorithm to solve the basic problem is what we will call the \emph{fallback solution}. The goal of this algorithm is to be a computationally quick and easy to implement and to attempt to minimize the total cost of the transaction at the same time.

In detail, assume we are given $M$ payment requests $J^* \subseteq \mathcal{P}$ to process. Expanding the cost function, we see that
$$
\text{Cost}(I,J^*,c,r) = \gamma(10 +148|I| + 34M +34(1-\delta_{c,0})) + r
$$
is clearly minimized when the transaction is change-free, has zero overpayment and utilizing the minimal number of inputs. Moreover, the number of inputs used is most significant as (assuming the transactions are good)
$$
\cost(I,J^*,c,r) < \cost(I', J^*, 0,0)
$$
for any $I, I',c,r$ with $|I| < |I'|$. In down to earth terms, this means that when trying to construct a transaction, a good transaction with possible non-zero change and overpayment is more desirable than a good change-free and zero overpayment transaction with more UTXO inputs. 

To this end, recalling $\mathcal{U} = \{ u_1 \geq u_2 \geq \cdots \geq u_n \}$ is written in decreasing order, we define for $J \subset \mathcal{P}$ some collection of payment requests,
\begin{equation}
    \text{opt}(J) := \text{min} \{ k \,\, | \,\, (\{u_j\}_1^k, J, c, 0) \, \text{is a good transaction for some} \, c \}.
\end{equation}
In other words, $\text{opt}(J)$ is the minimal number of UTXOs needed to produce \emph{some} good transaction for the pay requests $J$. The fallback solution is then the transaction taking the top $\text{opt}(J^*)$ UTXOs as inputs.
\begin{defx}
The Fallback Solution to process the payment requests $J^* \subset \mathcal{P}$ is the transaction $(I, J^*, c,0)$ where
\begin{enumerate}
    \item $I = \{ u_1, u_2, \dots, u_{\text{opt}(J)} \}$
    \item $c = \sum_{j=1}^{\text{opt}(J)}u_j - \sum_{p \in J^*}p - s(I,J^*,c,0)\gamma$.
\end{enumerate}
\end{defx}

\subsection{Basic Problem - Knapsack Solution}

The main goal of the knapsack solution to a basic problem is to produce a change-free transaction to save on the cost of having a change output. 

Let $J^* \subset \mathcal{P}$ be a collection of $M$ payment requests which must be processed. In this case we are seeking a minimal cost change-free transaction, i.e. we are attempting to solve the optimization problem
\begin{equation*} \label{basicprob}
    \underset{I,r}{\text{arg min}} \, \, \text{Cost}(I, J^*, 0, r)
\end{equation*}
subject to:
$$
(I, J^*, 0, r) \,\, \text{is a good transaction.}
$$

For programming purposes, this problem should be translated into an \emph{integer linear program}, or rather in our case, a \emph{binary linear program}. Recall, such a problem is one of the form
\begin{equation}
    \underset{x}{\text{min}} \, c^Tx
\end{equation}
subject to:
$$
Ax \leq b
$$
$$
x \in \{0, 1\}^n
$$
for $x,c \in \mathbb{R}^n$, $b \in \mathbb{R}^m$ and $A \in \mathbb{R}^{m \times n}$. We formulate our problem as one such now. 
Let $x_1, x_2, \dots, x_n$ denote the decision variables, i.e.
$$
x_j = \begin{cases}
        1 & \text{if UTXO} \,\, u_j \,\, \text{is included in the transaction} \\
        0 & \text{otherwise}
        \end{cases}
$$

From above, note that the optimal solution $(I,J^*, 0, r)$ to our problem must satisfy $|I| = \text{opt}(J^*)$. Hence, minimizing $\cost(I,J^*,0,r)$ is in fact the same as minimizing the overpayment $r$ along with the extra constraint $|I| = \text{opt}(J^*)$. Moreover, the overpayment $r$ is then immediately written in terms of the decision variables $x_i$,
$$
r = \sum_{j=1}^n x_ju_j - \sum_{j=1}^M p_j - \text{size}(\text{opt}(J^*), M, 0)\gamma
$$
Recall the function $\text{size}$ from above. With these observations, our problem as an binary linear program is:
\begin{equation} \label{knap_prob}
    \underset{x_i}{\text{arg min}} \left( \sum_{j=1}^n x_ju_j - \sum_{j=1}^M p_j - \text{size}(\text{opt}(J^*), M, 0)\gamma \right)
\end{equation}
subject to:
\begin{enumerate}
    \item The number of inputs used must be optimal:
    $$ \sum_{j=1}^n x_j = \text{opt}(J^*) $$
    \item The transaction must be good:
    $$
    \sum_{j=1}^n x_ju_j - \sum_{j=1}^M p_j - \text{size}(\text{opt}(J^*), M, 0)\gamma \geq 0
    $$
    $$
    \sum_{j=1}^n x_ju_j - \sum_{j=1}^M p_j - \text{size}(\text{opt}(J^*), M, 0)\gamma \leq h
    $$
    \item Each $x_j$ is binary:
    $$
    x_j \in \{0, 1\} \hspace{1cm} \text{for all} \,\, j=1,\dots,n
    $$
\end{enumerate}

\begin{defx}
The Knapsack Solution to process the payment requests $J^* \subset \mathcal{P}$ with UTXO pool $\mathcal{U}$ is the transaction $(I,J^*,0,r)$ which solves the optimization problem (\ref{knap_prob}) i.e.
\begin{enumerate}
    \item $I = \{ u_j \in \mathcal{U} \, | \, x_j =1 \} \subset \mathcal{U}$
    \item $r= \sum_{j=1}^n x_ju_j - \sum_{j=1}^M p_j - \text{size}(\text{opt}(J^*), M, 0)\gamma$
\end{enumerate}
\end{defx}

In practice, after some time period we usually cut off the algorithms search for the optimal solution to \ref{knap_prob} and accept a feasible solution. Depending of the size of the various parameters, search times for the optimal solutions can be considered too long and quickly finding \emph{some} feasible solution is acceptable.

\subsection{Basic Problem - Knapsack with Leverage Solution} The idea of the standard knapsack algorithm above is to try and find a minimal cost, good, change-free transaction which processes a specified collection of pay requests $J^*$ by solving a binary linear program.

The attempt at finding any feasible solution to the knapsack problem may fail for several reasons (e.g. there is no solution or the algorithm fails to produce one in a certain allotted time period). The first full procedure outlined below, attempting to solve the full problem, first trys to find a knapsack solution, and if it fails to then utilze the fallback solution.

The idea for the leverage solution is to not immediately use the fallback solution but try and exploit the fact that once the standard knapsack algorithm fails, it is known that there will be a change output in the transaction. The leverage solution trys to produce such a transaction for which the change output is useful as a future UTXO. That is to say, the leverage solution attempts to construct two transactions, one processing the current pay requests $J^*$ and another processing some other set $J_2$, so that the change output of the first transaction fits precisely into the second making it change-free.

More precisely, the leverage solution attempts to find $I_1, I_2 \subset \mathcal{U}$, $J_2 \subset \mathcal{U}\setminus J^*$, $c_1, r_1, r_2 \in \mathbb{R}_{\geq 0}$ so that:
\begin{enumerate}
    \item $(I_1, J^*, c_1, r_1)$ is a good transaction processing payment requests $J^*$
    \item $(I_2 \cup \{c_1\}, J_2, 0, r_2)$ is a good change-free transaction processing payment requests $J_2$.
\end{enumerate}

One should note that in practice the change output $c_1$ cannot be used as input UTXO to process the second transaction until it is mined and a part of the blockchain. So in that sense, the pay requests $J_2$ could be representative of lower priority pay requests (they will be processed in a block after that of the first transaction). A possible small alteration to the algorithm below would be to restrict the search for an optimal $J_2$ to a certain subset of $\mathcal{P}$ (other than $\mathcal{P} \setminus J^*$) that may be more representative of these``low priority transactions".

Let us formulate the leverage solution as a binary linear program. Our decision variables are:
$$
x_{i,j} = \begin{cases}
        1 & \text{if UTXO} \,\, u_j \,\, \text{is included in transaction} \,\, i \\
        0 & \text{otherwise}
        \end{cases}
$$
$$
y_j = \begin{cases}
        1 & \text{if pay request} \,\, p_j \,\, \text{is included in transaction} \,\, 2 \\
        0 & \text{otherwise}
        \end{cases}
$$

For simplicity let us assume that the first transaction will always be responsible for processing the first $M$ payment requests, $J^* = \{p_1, \dots, p_M\}$. This allows us to focus on the decision variables $y_{M+1}, y_{M+2}, \dots, y_m$.

We then consider the following binary linear program:
\begin{equation}
    \underset{x_{1,j}, x_{2,j}, y_j}{\text{arg min}} \,\, \sum_{j=1}^n x_{2,j}
\end{equation}
subject to:
\begin{enumerate}
    \item \label{levcons1}Use a UTXO at most once:
    $$
    x_{1,j} + x_{2,j} \leq 1, \hspace{1cm} \text{for all} \,\, j=1,\dots,n
    $$
    \item \label{levcons2} Use the optimal number of inputs for transaction 1:
    $$
    \sum_{j=1}^n x_{1,j} = \text{opt}(J^*)
    $$
    \item \label{levcons3} Process a certain number of additional pay requests:
    $$
    M_1 \leq \sum_{j=M+1}^m y_j \leq M_2
    $$
    \item \label{levcons4} Transaction 1 is valid with change:
    $$
    \sum_{j=1}^n x_{1,j}u_j \geq \sum_{p \in J^*}p + \text{size}(\text{opt}(J^*), M, 1)\gamma
    $$
    \item \label{levcons5} Transaction 2 ``needs a change UTXO":
    $$
    \sum_{j=1}^n x_{2,j}u_j \leq \sum_{j=M+1}^m y_jp_j + \text{size}\left(1 + \sum_{j=1}^n x_{2,j}, \sum_{j=M+1}^m y_j, 0\right ) \gamma
    $$
    \item \label{levcons6} Enforce change output is correct (1 of 2):
    \begin{equation}
        \begin{split}
            & \sum_{j=1}^n x_{1,j}u_j - \sum_{p \in J^*}p - \text{size}(\text{opt}(J^*), M, 1)\gamma \geq \\
            & \sum_{j=M+1}^m y_jp_j + \text{size}\left(1 + \sum_{j=1}^n x_{2,j}, \sum_{j=M+1}^m y_j, 0\right ) \gamma - \sum_{j=1}^n x_{2,j}u_j
        \end{split}
    \end{equation}
    \item \label{levcons7} Enforce change output is correct (2 of 2)
    \begin{equation}
        \begin{split}
            & \sum_{j=1}^n x_{1,j}u_j - \sum_{p \in J^*}p - \text{size}(\text{opt}(J^*), M, 1)\gamma \leq \\
            & \sum_{j=M+1}^m y_jp_j + \text{size}\left(1 + \sum_{j=1}^n x_{2,j}, \sum_{j=M+1}^m y_j, 0\right ) \gamma - \sum_{j=1}^n x_{2,j}u_j + \beta h.
        \end{split}
    \end{equation}
\end{enumerate}
Note that for brevity we have used the size function and this does not break the linearity of the constraints as size itself is affine linear.

Some more explanation of the objective function and the constraints may be needed:
\begin{itemize}
    \item Objective Function: This function counts one fewer than the number of inputs that will be used in the second transaction (as one additional input will be utilized as the change from the first transaction). The binary linear program attempts to minimize this number so that the total cost of the second transaction is minimized.
    \item Constraint (\ref{levcons1}): This is to simply guarantee that a single UTXO is not used in both transactions.
    \item Constraint (\ref{levcons2}): This is to enforce the cost of the first transaction is minimal (as here we are assuming there will be a change output).
    \item Constraint (\ref{levcons3}): The number of pay requests processed in transaction 2 should be between two user set parameters $M_1$ and $M_2$. These parameters are set at the users discretion to help aid in addressing other goals in mind for coin selection.
    \item Constraint (\ref{levcons4}): Transaction 1 must be a valid transaction, i.e. the inputs must be able to cover the cost of the outputs and transaction fee.
    \item Constraint (\ref{levcons5}): The total input amounts for transaction 2 must fall short of what is needed to make the transaction valid. The gap will be filled by the change output from transaction 1.
    \item Constraint (\ref{levcons6}): First of two conditions enforcing that the change output from transaction 1 will match up with what is needed to make a change-free good transaction 2.
    \item Constraint (\ref{levcons7}): Second of these two conditions. We remark that we are not looking for a perfect match coming from the transaction 1 change $c_1$, just one that allows for the overpayment $r_2$ of transaction 2 to be less than the product of the make-change threshold $h$ and a parameter we will refer to as the \emph{leverage boost factor} $\beta$. $\beta$ is to take values in $[0,1]$ and should intuitively represent how ``greedy" the algorithm is being when it comes to reducing the overpayment amount. We will expand on $\beta$ and how it's chosen in Section \ref{sim_results}.
\end{itemize}

\begin{defx}
The Knapsack with Leverage solution to process payment requests $J^* \subset \mathcal{P}$ is the pair of transactions $(I_1, J^*, c_1, 0)$ and $(I_2, J_2, 0, r_2)$ where:
\begin{enumerate}
    \item $I_1 = \{ u_j \in \mathcal{U} \, | \, x_{1,j} = 1 \}$
    \item $I_2 = \{ u_j \in \mathcal{U} \, | \, x_{2,j} = 1 \} \cup \{ c_1 \}$
    \item $J_2 = \{ p_j \in \mathcal{P}\setminus J^* \, | \, y_j = 1 \}$
    \item $c_1 = \sum_{j=M+1}^m y_jp_j + \text{size}\left(1 + \sum_{j=1}^n x_{2,j}, \sum_{j=M+1}^m y_j, 0\right ) \gamma - \sum_{j=1}^n x_{2,j}u_j$ (The change UTXO created by processing the first transaction which will allow transaction 2 to be change free).
    \item $r_2 = \sum_{j=1}^n x_{1,j}u_j - \sum_{p \in J^*}p - \text{size}(\text{Opt}(J^*), M, 1)\gamma - c_1$ (The slight overpayment for transaction 2, chosen to be less than or equal to $h$).
\end{enumerate}
\end{defx}

\section{The Full Problem and Solutions}
\label{full_soln}

Recall that by the \emph{full} problem we mean the process of constructing transactions $T_1 = (I_1, J_1, c_1, r_1), \dots, T_K = (I_K, J_K, c_K, r_K)$ that as a whole will process \emph{at least} all of some fixed sub collection of payment requests $\mathcal{P}' \subset \mathcal{P}$, i.e. $\cup J_k \supseteq \mathcal{P}'$ with the $J_k$ pairwise disjoint. With combining the three algorithms for solving the basic problem above, we construct two full algorithms which aim to solve the full problem.

One subtlety that should be addressed, but shouldn't cause confusion is the fact that the UTXO set can change. In particular, all pay requests, after they are processed are transferred into a new UTXO. For a wallet provider, when a change output is created, it is returned to their UTXO set and thus can be used in a future transaction. Hence, in the transactions above, the set of inputs $I_k$ strictly speaking need not be subsets of the original UTXO set $\mathcal{U}$, but rather could also contain new UTXOs generated by processed pay requests at an earlier time. This of course, is at the heart of the Knapsack Leverage algorithm.

\subsection{Full Problem - Knapsack Solution} \label{fpks}

The solution is comprised of performing a number of repetitive iterations of the same sequence of steps, after which the UTXO pool and payment request pool is updated.

Let $\mathcal{P}' \subset \mathcal{P}$ be a sub collection of payment requests with which to process. Let $\mathcal{U}_k, \mathcal{P}_k, \mathcal{P}'_k$ be the updated UTXO and payment request pools along with the updated current payment requests to be processed respectively after iteration $k$. Thus, $\mathcal{U}_0 = \mathcal{U}, \mathcal{P}_0 = \mathcal{P}$ and $\mathcal{P}'_0 = \mathcal{P}'$.

The Knapsack solution to the full problem is then: after completing iteration $k-1$, iteration $k$ is 
\begin{enumerate}
    \item Consider $J_{k}^* \subset \mathcal{P}'_{k-1}$ the top $M$ remaining payment requests according to \emph{urgency}.
    \item \label{knapattempt} Attempt to solve the basic problem of processing $J_k^*$ with UTXO pool $\mathcal{U}_{k-1}$ via a knapsack solution. If solution $(I_k, J^*_k, 0, r_k)$ is found, payment requests $J_k^*$ have been processed. Update UTXO pool $\mathcal{U}_{k} = \mathcal{U}_{k-1} \setminus I_k$, payment request pool $\mathcal{P}_{k} = \mathcal{P}_{k-1} \setminus J^*_k$ and current payment requests $\mathcal{P}'_{k} = \mathcal{P}'_{k-1} \setminus J^*_k$
    \item If (\ref{knapattempt}) fails, resort to utilizing the fallback solution $(I_k, J^*_k, c_k, r_k)$. Update UTXO and payment request pool $\mathcal{U}_{k} = (\mathcal{U}_{k-1} \setminus I_k) \cup \{c_k\}$, $\mathcal{P}_{k} = \mathcal{P}_{k-1} \setminus J^*_k$ as well as current payment requests $\mathcal{P}'_{k} = \mathcal{P}'_{k-1} \setminus J^*_k$
    \item Continue onto next iteration until all of $\mathcal{P}'$ is processed.
\end{enumerate}

\subsection{Full Problem - Knapsack with Leverage Solution} This solution attempts to improve on the standard Knapsack solution of Section (\ref{fpks}) in reducing the total cost to process all payment requests $\mathcal{P}$.

As above, the solution is also comprised of performing a number of repetitive iterations of the same sequence of steps, after which the UTXO pool and payment request pool is updated.

Using the same notation for $\mathcal{U}_k, \mathcal{P}_k$ and $\mathcal{P}'_k$ as above, the knapsack with Leverage solution to the full problem is then: after completing iteration $k-1$, iteration $k$ is 

\begin{enumerate}
    \item Consider $J_{k}^* \subset \mathcal{P}_{k-1}$ the top $M$ remaining payment requests according to \emph{urgency}.
    \item \label{knapattemptLEV} Attempt to solve the basic problem of processing $J_k^*$ with UTXO pool $\mathcal{U}_{k-1}$ via a knapsack solution. If solution $(I_k, J^*_k, 0, r_k)$ is found, payment requests $J_k^*$ have been processed. Update UTXO pool $\mathcal{U}_{k} = \mathcal{U}_{k-1} \setminus I_k$, payment request pool $\mathcal{P}_{k} = \mathcal{P}_{k-1} \setminus J^*_k$ and current payment requests $\mathcal{P}'_{k} = \mathcal{P}'_{k-1} \setminus J^*_k$
    \item \label{levattempt} If (\ref{knapattemptLEV}) fails then attempt to solve the basic problem using the knapsack with leverage approach. If a solution is found producing two transactions $(I_k, J^*_k, c_k, 0)$, $(I_k', J_k', 0, r_k')$ then the payment requests $J^*_k \cup J_k'$ have been processed. Update UTXO and payment request pool as $\mathcal{U}_{k} = \mathcal{U}_{k-1} \setminus I_k, \mathcal{P}_{k} = \mathcal{P}_{k-1} \setminus (J^*_k \cup J'_k)$ and current payment requests $\mathcal{P}'_{k} = \mathcal{P}'_{k-1} \setminus (J^*_k \cup J'_k)$.
    \item If (\ref{levattempt}) fails, resort to utilizing the fallback solution $(I_k, J^*_k, c_k, r_k)$. Update UTXO and payment request pool $\mathcal{U}_{k} = (\mathcal{U}_{k-1} \setminus I_k) \cup \{c_k\}$, $\mathcal{P}_{k} = \mathcal{P}_{k-1} \setminus J^*_k$ as well as current payment requests $\mathcal{P}'_{k} = \mathcal{P}'_{k-1} \setminus J^*_k$
    \item Continue onto next iteration until all of $\mathcal{P}'$ is processed.
\end{enumerate}

\section{Simulation Results}
\label{sim_results}

Here we present results of several simulations run testing and comparing the above two appraoches. The data used in the simulations were obtained from the actual UTXO set on October 1st, 2019 and the payment requests were generated by sampling credit card transaction data taken from the IEEE-CIS Fraud Detection Kaggle competition \cite{kaggle_data}.

The simulations done used the following parameters:
\begin{enumerate}
    \item A UTXO pool of $2,500$ UTXOs, generated by a random sample from the UTXO set of October 1st, 2019.
    \item A payment request pool of $250$ payment requests, sampled from credit card transaction data \cite{kaggle_data}. Transactions worth less than the Satoshi equivalent of 4004 are discarded. 
    \item The fee per byte rates $\gamma$ tested are $\gamma = 22, 60, 200, 400, 900$.
    \item The parameters $M_1, M_2$ are deliberately chosen to agree $M_1 = M_2$ and that their common value $M$ is one of $2,3,5, 10$. The reason for this choice is to be able to more accurately attribute the savings to the Leverage algorithm, over simply the fact that more payment requests per transactions could be processed.  We remark that increasing the difference $M_2 - M_1 > 0 $ inherently increases the likelihood that leverage algorithm succeeds, thus even greater savings is expected in that case.
    \item The dollar values below assume $1 \, \text{BTC} = \$ 8,582$ unless otherwise indicated.
    \item The leverage boost factor $\beta$ has been chosen based on experimentation. A value of $\beta < 1$ is sometimes required due to the unpredictability of the various overpayment amounts. A smaller value of $\beta$ is used to help produce more savings for the leverage technique. The precise values used are indicated in the Table \ref{boost_factor_table}.
\end{enumerate}

\begin{center}
    \input{beta_table.tex}

\end{center}

With the above understood, the simulations are carried out in the following way: For each choice for the pair $(\gamma, M)$, a random sample of $2,500$ UTXOs and $250$ payment requests are drawn and each full algorithm (Leverage and No Leverage) is run for \emph{5} iterations. New samples of UTXOs and payment requests are then drawn and $5$ iterations are run. This process is repeated \emph{10} times, guaranteeing that at least $50M$ payment requests are processed. The Tables \ref{nolev_sum}, \ref{lev_sum} and \ref{savings_sum} summarize the results.

\newpage

\begin{center}
    \input{noLev_summary.tex}

\end{center}

\begin{center}
    \input{lev_summary.tex}

\end{center}

\newpage

\begin{center}
    \input{savings_summary.tex}

\end{center}

As we see, across the various chosen parameters we saw a savings of about $1-2$\% of cost per payment request. Depending on market conditions, this seemingly small amount can add up quickly.

There are approximately $300,000$ confirmed Bitcoin transactions per day. For a fictitious Bitcoin wallet provider/exchange that represents, say $\sim 2\%$ of all transactions, we can estimate that such an exchange would have approximately $12,000 - 15,000$ payment requests to process each day. 

It is an assumption by the author that for such a fictitious wallet provider, utilizing a UTXO and payment request pool size of $2,500$ and $250$ respectively, is reasonable. Although the author was unable to find data regarding UTXO pool sizes, for entities with access to a large UTXO pool, it may be reasonable to imagine first sampling a sub-pool to use for coin selection (for speed considerations for example). In that scenario the results shown here may prove useful.

For such a wallet provider we estimate the following savings depending on market conditions: Tables \ref{at_scale01}, \ref{at_scale03}, and \ref{at_scale04} assume the value of a Bitcoin is similar to that of October 2019 ($\$ 8,582$) while Table \ref{at_scale02} estimates savings in \emph{extreme} market conditions akin to January 2018.
\newpage

\begin{center}
    \input{at_scale01.tex}

\end{center}

\begin{center}
    \input{at_scale03.tex}

\end{center}

\begin{center}
    \input{at_scale04.tex}

\end{center}

\begin{center}
    \input{at_scale02.tex}

\end{center}

\vspace{-.75in}

 While the raw numbers above may be seen as optimistic, it is an illustration at the potential in cost savings, especially in extreme market conditions as in January 2018. Of course, it is not clear if and when conditions will be that extreme again. Nevertheless, even in relatively mild market conditions, Tables \ref{at_scale01}, \ref{at_scale03} and \ref{at_scale04} shows the utility of our leverage technique.

In practice, a knapsack algorithm may be part of a collection of techniques used in sequence in a entity's coin selection procedure. Our new leverage technique should naturally fit in as a replacement for the knapsack step. Furthermore, we recognize that cost savings is not the only goal in mind when considering a coin selection procedure. To this end, one chooses the values of the various parameters such as $M_1, M_2$ and $\beta$ based on these goals.

The code used to run these simulations is publicly available on GitHub \cite{GitHub}. All simulations were written in python and the various optimization problems were solved with the aid of the Python library PuLP.

\bibliographystyle{plain}
\bibliography{bibfile}
\nocite{*}
\end{document}

%% file: beta_table.tex
\begin{table}[]
    \centering
    \begin{tabular}{llr}
    \toprule
        &    &  \text{Leverage Boost Factor} $\beta$ \\
    $\gamma$ \,\, (\text{Satoshi Per Byte}) & $M$ &              \\
    \midrule
    22  & 2  &         0.94 \\
        & 3  &         0.96 \\
        & 5  &         1.00 \\
        & 10 &         1.00 \\
    60  & 2  &         0.78 \\
        & 3  &         0.96 \\
        & 5  &         0.94 \\
        & 10 &         0.98 \\
    200 & 2  &         0.54 \\
        & 3  &         0.64 \\
        & 5  &         0.84 \\
        & 10 &         0.96 \\
    400 & 2  &         0.52 \\
        & 3  &         0.52 \\
        & 5  &         0.66 \\
        & 10 &         0.86 \\
    900 & 2  &         0.22 \\
        & 3  &         0.44 \\
        & 5  &         0.64 \\
        & 10 &         0.82 \\
    \bottomrule
    \end{tabular}
    \caption{Choice of Leverage Boost Factor parameter $\beta$ used in simulations}
    \label{boost_factor_table}
\end{table}

%% file: noLev_summary.tex
\begin{table*}[]
    \centering
    \resizebox{\textwidth}{!}{
    \begin{tabular}{llrrrrr}
    \toprule
        &    &  Fallback &  Knapsack &  Leverage &  Payment Requests &  Cost per Payment \\
    $\gamma$ & $M$ &  Success Rate                      &       Success Rate                 &      Success Rate                  &                  Processed       &            Request (in USD)                              \\
    \midrule
    22  & 2  &                   0.94 &                   0.06 &                    0.0 &                      100 &                                 \$0.244890 \\
        & 3  &                   0.96 &                   0.04 &                    0.0 &                      150 &                                 \$0.184722 \\
        & 5  &                   1.00 &                   0.00 &                    0.0 &                      250 &                                 \$0.136694 \\
        & 10 &                   1.00 &                   0.00 &                    0.0 &                      500 &                                 \$0.100444 \\
    60  & 2  &                   0.78 &                   0.22 &                    0.0 &                      100 &                                 \$0.657344 \\
        & 3  &                   0.96 &                   0.04 &                    0.0 &                      150 &                                 \$0.503280 \\
        & 5  &                   0.94 &                   0.06 &                    0.0 &                      250 &                                 \$0.372113 \\
        & 10 &                   0.98 &                   0.02 &                    0.0 &                      500 &                                 \$0.273701 \\
    200 & 2  &                   0.54 &                   0.46 &                    0.0 &                      100 &                                 \$2.159708 \\
        & 3  &                   0.64 &                   0.36 &                    0.0 &                      150 &                                 \$1.645127 \\
        & 5  &                   0.84 &                   0.16 &                    0.0 &                      250 &                                 \$1.232979 \\
        & 10 &                   0.96 &                   0.04 &                    0.0 &                      500 &                                 \$0.911609 \\
    400 & 2  &                   0.52 &                   0.48 &                    0.0 &                      100 &                                 \$4.281211 \\
        & 3  &                   0.52 &                   0.48 &                    0.0 &                      150 &                                 \$3.244493 \\
        & 5  &                   0.66 &                   0.34 &                    0.0 &                      250 &                                 \$2.441034 \\
        & 10 &                   0.86 &                   0.14 &                    0.0 &                      500 &                                 \$1.818567 \\
    900 & 2  &                   0.22 &                   0.78 &                    0.0 &                      100 &                                 \$9.273401 \\
        & 3  &                   0.44 &                   0.56 &                    0.0 &                      150 &                                 \$7.261839 \\
        & 5  &                   0.64 &                   0.36 &                    0.0 &                      250 &                                 \$5.493654 \\
        & 10 &                   0.82 &                   0.18 &                    0.0 &                      500 &                                 \$4.085761 \\
    \bottomrule
    \end{tabular}}
    \caption{Simulation results utilizing standard Knapsack approach.}
    \label{nolev_sum}
\end{table*}

%% file: lev_summary.tex
\begin{table}[]
    \centering
    \resizebox{\textwidth}{!}{
    \begin{tabular}{llrrrrr}
    \toprule
       &    &  Fallback &  Knapsack &  Leverage &  Payment Requests &  Cost per Payment \\
    $\gamma$ & $M$ &  Success Rate                      &       Success Rate                 &      Success Rate                  &                  Processed       &            Request (in USD)                              \\
    \midrule
    22  & 2  &                   0.52 &                   0.04 &                   0.44 &                      144 &                                 \$0.239597 \\
        & 3  &                   0.54 &                   0.02 &                   0.44 &                      216 &                                 \$0.181566 \\
        & 5  &                   0.58 &                   0.00 &                   0.42 &                      355 &                                 \$0.134602 \\
        & 10 &                   0.66 &                   0.00 &                   0.34 &                      670 &                                 \$0.099785 \\
    60  & 2  &                   0.20 &                   0.20 &                   0.60 &                      160 &                                 \$0.641769 \\
        & 3  &                   \$0.36 &                   0.06 &                   0.58 &                      237 &                                 \$0.492071 \\
        & 5  &                   0.38 &                   0.08 &                   0.54 &                      385 &                                 \$0.365785 \\
        & 10 &                   0.48 &                   0.02 &                   0.50 &                      750 &                                 \$0.270931 \\
    200 & 2  &                   0.00 &                   0.40 &                   0.60 &                      160 &                                 \$2.107687 \\
        & 3  &                   0.24 &                   0.26 &                   0.50 &                      225 &                                 \$1.615732 \\
        & 5  &                   0.32 &                   0.16 &                   0.52 &                      380 &                                 \$1.208484 \\
        & 10 &                   0.44 &                   0.04 &                   0.52 &                      760 &                                 \$0.899072 \\
    400 & 2  &                   0.02 &                   0.52 &                   0.46 &                      146 &                                 \$4.174872 \\
        & 3  &                   0.10 &                   0.48 &                   0.42 &                      213 &                                 \$3.182614 \\
        & 5  &                   0.24 &                   0.36 &                   0.40 &                      350 &                                 \$2.400355 \\
        & 10 &                   0.46 &                   0.06 &                   0.48 &                      740 &                                 \$1.804518 \\
    900 & 2  &                   0.02 &                   0.78 &                   0.20 &                      120 &                                 \$9.221179 \\
        & 3  &                   0.14 &                   0.68 &                   0.18 &                      177 &                                 \$7.142914 \\
        & 5  &                   0.26 &                   0.34 &                   0.40 &                      350 &                                 \$5.423762 \\
        & 10 &                   0.46 &                   0.22 &                   0.32 &                      660 &                                 \$4.036800 \\
    \bottomrule
    \end{tabular}}
    \caption{Simulation results utilizing new Leverage technique.}
    \label{lev_sum}
\end{table}

%% file: savings_summary.tex
\begin{table}[]
    \centering
    \begin{tabular}{llrr}
    \toprule
        &    &  \% Savings Per Payment&  Savings Per Payment \\
    $\gamma$ & $M$ &                    Request        &  Request (in USD)                 \\
    \midrule
    22  & 2  &                   2.161472 &                 \$0.005293 \\
        & 3  &                   1.708508 &                 \$0.003156 \\
        & 5  &                   1.530141 &                 \$0.002092 \\
        & 10 &                   0.655868 &                 \$0.000659 \\
    60  & 2  &                   2.369357 &                 \$0.015575 \\
        & 3  &                   2.227315 &                 \$0.011210 \\
        & 5  &                   1.700473 &                 \$0.006328 \\
        & 10 &                   1.012060 &                 \$0.002770 \\
    200 & 2  &                   2.408727 &                \$0.052021 \\
        & 3  &                   1.786809 &                 \$0.029395 \\
        & 5  &                   1.986702 &                \$0.024496 \\
        & 10 &                   1.375276 &                \$0.012537 \\
    400 & 2  &                   2.483857 &                \$0.106339 \\
        & 3  &                   1.907207 &                \$0.061879 \\
        & 5  &                   1.666475 &                \$0.040679 \\
        & 10 &                   0.772521 &                \$0.014049 \\
    900 & 2  &                   0.563130 &                \$0.052221 \\
        & 3  &                   1.637678 &                \$0.118926 \\
        & 5  &                   1.272223 &                \$0.069892 \\
        & 10 &                   1.198339 &                \$0.048961 \\
    \bottomrule
    \end{tabular}
    \caption{A summary table comparing the Leverage technique to the standard Knapsack approach.}
    \label{savings_sum}
\end{table}

%% file: at_scale01.tex
\begin{table}[]
    \centering
    \resizebox{\textwidth}{!}{
    \begin{tabular}{llrrll}
    \toprule
       &    &  \% Savings Per &  Savings Per  &   Savings Per Day &     Savings Per Month \\
    $\gamma$ & $M$ &             Payment Request               &                   Payment Request       &                   &                       \\
    \midrule
    22 & 2  &                   2.161472 &                 \$0.005293 &   \$63.52 -  \$79.4 &  \$1,905.56 -  \$2,381.95 \\
       & 3  &                   1.708508 &                 \$0.003156 &  \$37.87 -  \$47.34 &   \$1,136.16 -  \$1,420.2 \\
       & 5  &                   1.530141 &                 \$0.002092 &   \$25.1 -  \$31.37 &    \$752.98 -  \$941.23 \\
       & 10 &                   0.655868 &                 \$0.000659 &    \$7.91 -  \$9.88 &    \$237.16 -  \$296.45 \\
    \bottomrule
    \end{tabular}}
    \caption{A glance at savings approximating market conditions from October 2019 with one BTC = $\$8,582$ and $\gamma = 22$.}
    \label{at_scale01}
\end{table}

%% file: at_scale03.tex
\begin{table}[]
    \centering
    \resizebox{\textwidth}{!}{
    \begin{tabular}{llrrll}
    \toprule
       &    &  \% Savings Per&  Savings Per&     Savings Per Day &     Savings Per Month \\
    $\gamma$ & $M$ &          Payment Request                  &             Payment Request              &                     &                       \\
    \midrule
    60 & 2  &                   2.369357 &                 \$0.015575 &   \$186.9 -  \$233.62 &  \$5,606.93 - \$7,008.67 \\
       & 3  &                   2.227315 &                 \$0.011210 &  \$134.52 -  \$168.14 &  \$4,035.47 -  \$5,044.34 \\
       & 5  &                   1.700473 &                 \$0.006328 &    \$75.93 -  \$94.92 &  \$2,277.97 -  \$2,847.46 \\
       & 10 &                   1.012060 &                 \$0.002770 &    \$33.24 -  \$41.55 &    \$997.2 -  \$1,246.51 \\
    \bottomrule
    \end{tabular}}
    \caption{A glance at savings approximating market conditions from October 2019 with one BTC = $\$8,582$ and $\gamma = 60$.}
    \label{at_scale03}
\end{table}

%% file: at_scale04.tex
\begin{table}[]
    \centering
    \resizebox{\textwidth}{!}{
    \begin{tabular}{llrrll}
    \toprule
        &    &  \% Savings Per&  Savings Per &     Savings Per Day &       Savings Per Month \\
   $\gamma$ & $M$ &            Payment Request                &    Payment Request                      &                     &                         \\
    \midrule
    200 & 2  &                   2.408727 &                 \$0.052021 &  \$624.26 -  \$780.32 &  \$18,727.73 -  \$23,409.66 \\
        & 3  &                   1.786809 &                 \$0.029395 &  \$352.74 -  \$440.93 &   \$10,582.3 -  \$13,227.87 \\
        & 5  &                   1.986702 &                 \$0.024496 &  \$293.95 -  \$367.43 &   \$8,818.42 -  \$11,023.03 \\
        & 10 &                   1.375276 &                 \$0.012537 &  \$150.45 -  \$188.06 &    \$4,513.37 -  \$5,641.71 \\
    \bottomrule
    \end{tabular}}
    \caption{A glance at savings approximating market conditions from October 2019 with one BTC = $\$8,582$ and $\gamma = 200$.}
    \label{at_scale04}
\end{table}

%% file: at_scale02.tex
\begin{table}[]
    \centering
    \resizebox{\textwidth}{!}{
    \begin{tabular}{llrrll}
    \toprule
        &    &  \% Savings Per &  Savings Per &       Savings Per Day &        Savings Per Month \\
    $\gamma$ & $M$ &                 Payment Request            &   Payment Request                        &                       &                          \\
    \midrule
    900 & 2  &                   0.563130 &                 \$0.104503 &  \$1,254.04 -  \$1,567.55 &   \$37,621.25 -  \$47,026.57 \\
        & 3  &                   1.637678 &                 \$0.237990 &  \$2,855.88 -  \$3,569.84 &  \$85,676.28 -  \$107,095.35 \\
        & 5  &                   1.272223 &                 \$0.139864 &  \$1,678.37 -  \$2,097.97 &     \$50,351.2 -  \$62,939.0 \\
        & 10 &                   1.198339 &                 \$0.097980 &  \$1,175.76 -  \$1,469.69 &   \$35,272.66 -  \$44,090.82 \\
    \bottomrule
    \end{tabular}}
    \caption{A glance at savings approximating \emph{extreme} market conditions from January 2018 with {\bf one BTC = $\$17,174$} and $\gamma = 900$.}
    \label{at_scale02}
\end{table}